# CHEMO-MECHANICAL CHARACTERIZATION OF HYDRATED CALCIUM-HYDROSILICATES WITH COUPLED RAMAN- AND NANOINDENTATION MEASUREMENTS

P. Stemmermann [1], K. Garbev [1*], B. Gasharova [2], G. Beuchle [1], M. Haist [3], T. Divoux [4,5]

[1]Institute for Technical Chemistry, Karlsruhe Institute of Technology, 76021 Karlsruhe, Germany

[2]Institute for Beam Physics and Technology, Karlsruhe Institute of Technology, 76021 Karlsruhe, Germany

[3]Institute of Building Material Science, Leibniz University Hannover, 30167 Hannover, Germany

[4]MultiScale Material Science for Energy and Environment, UMI 3466, CNRS-MIT, 77 Massachusetts Avenue, Cambridge, Massachusetts 02139, USA

[5]Department of Civil and Environmental Engineering, MIT, Cambridge, MA 02139

Keywords: *hydraulic calcium –hydrosilicate, nanoindentation, Raman spectroscopy*

**Highlights**

- Study of the hydration of a novel isochemical cementitious material with a low $CO_2$ footprint.

- Lateral distribution and mechanical properties of single constituents in hardened cement paste studied by nanoindentation and Raman micro-spectroscopy.

- Excellent assignment of the mechanical properties to specific high density and low-density hydrate phases by combined nanoindentation and Raman micro-spectroscopy.

- Identification of Raman active markers for open and closed porosity in cementitious matrices.

---

*Corresponding author email: krassimir.garbev@kit.edu

**ABSTRACT**

Celitement is a new type of cement that is based on hydraulic calcium-hydrosilicate (hCHS). It is produced by mechanochemical activation of Calcium-Silicate-Hydrates (C-S-H) in a grinding process, whose intensity determines the reaction rate of celitement. Due to the lack of typical clinker minerals, its $CaO/SiO_2$ (C/S) ratio can be tuned between one and two. As a result, Celitement possesses a potential for minimizing the ratio C/S from above 3 in OPC down to 1, which significantly reduces the amount of $CO_2$ released during processing. The reaction kinetics of hCHS differs from that of classical clinker phases due to the presence of highly reactive silicate species, which involve silanol groups instead of pure calcium silicates and aluminates and aluminoferrites. In contrast to Portland cement, no calcium hydroxide is formed during hydration, which otherwise regulates the Ca concentration. Without the buffering role of $Ca(OH)_2$ the concentration of the dissolved species $c(Ca^{2+})$ and $c(SiO_4^{4-})$ and the corresponding pH must be controlled to ensure a reproducible reaction. Pure hCHS reacts isochemically with water, resulting in a C-S-H phase with the same chemical composition as a single hydration product, with a homogeneous distribution of the main elements Ca and Si throughout the sample. Here we study via nanoindentation tests, the mechanical properties of two different types of hardened pastes made out of Celitement (C/S=1.28), with varying amounts of hCHS and variable water to cement ratio. Usually, nanoindentation measurements are coupled with an elemental mapping of the same region of interest such as EDX or WDX to obtain information on the elemental distribution, and hence on the phase composition at the locus of each indent. However, such an assignment does not work in the case of hCHS, given the isochemical nature of its hydraulic reaction. Here we couple nanoindentation grids with Raman mappings to link the nanoscale mechanical properties to individual microstructural components, yielding in-depth insight into the mechanics of the mineralogical phases constituting the hardened cement paste. We show that we can identify in hardened Celitement paste both fresh C-S-H with varying density, and C-S-H from the raw material using their specific Raman spectra, while simultaneously measuring their mechanical properties, i.e., hardness, indentation modulus and creep modulus. Albeit not suitable for phase identification, supplemental EDX measurements provide valuable information about the distribution of alkalis, thus further helping to understand the reaction pattern of hCHS.

**INTRODUCTION**

The microstructure of hardened cement paste plays a crucial role in the determination of the properties of mortar and concrete. It is mainly composed of nanometer-sized Calcium Silicate Hydrates (C-S-H) and $Ca(OH)_2$ as major products of the cement hydration, along with residual unhydrated constituents and additional hydrated phases such as ettringite, monosulphate, etc. C-S-H displays different densities associated with more or less gel porosity. The former is referred to as High Density (HD) C-S-H and the latter as Low Density (LD) C-S-H [1]. The evolution of the paste microstructure described in several multiscale models and summarized in [2,3], results in the densification of the C-S-H and the corresponding change of pore size distribution over time [4]. Such a process is strongly dependent on the bulk chemistry, the water to cement (w/c) ratio, the mechanism, and kinetics of C-S-H nucleation and growth.

Mechanical properties of pure cement phases are challenging to determine macroscopically since the influence of the microstructure (e.g., pores) cannot be separated from the properties of the individual phases. Techniques such as nanoindentation offer the possibility to determine the mechanical properties directly *in situ* [5,6,7,8,9] when combined with statistical methods such as clustering analysis. To unambiguously identify specific cement phases from nanoindentation grids, the local chemical composition is determined by wavelength dispersive X-ray fluorescence spectroscopy (WDXRF) [5]. However, due to the composite microstructure of cement (a) almost all nanoindents are influenced by several phases, (b) the chemical composition is averaged across phase boundaries, (c) the measured analytical information stems from different volumes, and (d) the chemical information does not contain any structural information so that, for example, polymorphic modifications such as polymorphs of $C_2S$, HD and LD C-S-H cannot be distinguished.

Raman micro-spectroscopy overcomes the drawbacks of WDXRF, by delivering precise mineralogical information at small scales (<1 μm). The Raman spectra of C-S-H phases with variable C/S ratio, which corresponds to different degrees of silicate polymerization have been well studied [10,11]. Building upon this knowledge, we determine Raman maps, taken over the same area as indentation grids, that we analyze either by statistical methods including principal component analysis (PCA) and clustering or by direct imaging of characteristic Raman bands. Besides, semi-quantitative studies can be conducted at the locus of each indent by least-squares refinements, given the Raman spectra of the single constituents are available.

In this work, isochemical cement denotes all types of cement, which exclusively form C-S-H by reaction with water. C-S-H makes up for about 50 % by weight of the matrix in concrete and is responsible for its strength. Therefore, in principle, replacing ordinary Portland cement (OPC) in concrete with isochemical cement bears a potential for considerable saving in cement use with a corresponding reduction of $CO_2$ emissions. C-S-H formed by hydration of OPC has molar calcium to silicon ratio (C/S) of approximately 1.7. As C-S-H represents a solid solution series with C/S ranging from about 0.6 to 1.7, the use of isochemical cements with a lower C/S ratio would further reduce $CO_2$ emissions related to the processing of limestone. Although a lot of crystalline calcium silicates and calcium silicate hydrates exist in this C/S range, none of them would give rise to reactive material with the same stoichiometry upon thermal processing. Therefore, isochemical cements can only be derived from amorphous materials, e.g., glass. This work focusses on a particular type of isochemical cement called "Celitement" [12], which consists of amorphous nanoparticles made of hydraulic calcium hydro-silicate (hCHS) that are derived mechanochemically from dried C-S-H precursors [13,14,15]. Binders based on hCHS can be produced in a wide range of C/S, which results in hydrated products composed out of C-S-H with variable C/S ratio, featuring a broad range of reaction kinetics, microstructures and ultimately material properties. Reliable use of isochemical cements requires to describe analytically, model, and eventually control the hydration reactions and their kinetics. Since both the starting materials and the products of the hydraulic reaction are amorphous while the kinetics is controlled, as in OPC, by reaction zones in the micro- to the nanometer range, analytical characterization using classical methods, in particular XRD alone, is not sufficient. Finally, the absence of aluminates and ferrites in hardened Celitement paste makes the analytical examination of the reaction products simpler compared to those prepared from OPC.

The focus of this work is the analysis of hydrated samples of isochemical cement combining (i) Raman micro-spectroscopy for structural phase identification, (ii) Energy-dispersive X-ray spectroscopy (EDX) for quantifying the elemental distribution, especially that of potassium and (iii) nanoindentation for the determination of the local mechanical properties. We also propose a comparison of the different evaluation methods of Raman mappings and their suitability for distinguishing the phase composition of hardened Celitement pastes. Our results provide a comprehensive picture of the products of the hydration reactions with a lateral resolution of a few micrometers, which leads us to propose some qualitative thoughts on the reaction mechanisms. In particular, we show that in contrast to OPC, hardened Celitement pastes are not pH buffered by other products such as portlandite.

The sensitivity of the reactions to changes in parameters (e.g., pH) clearly shows that for technical application, a stable formulation of Celitement has to include stabilizing additives, e.g., alternative buffer systems.

**METHODS**

*2.1. Samples.*

Celitement binder are synthesized by granulating and autoclaving industrial-grade slaked lime and quartz sand with a C/S ratio of 1.28. The hydrothermal treatment is performed at 190°C and at the corresponding steam pressure for 6 hours. As a result, 85-90 wt.% of C-S-H (of type C-S-H I[16]) is formed together with unreacted raw materials. In a subsequent step, this precursor is ground to produce a hydraulically reactive cementitious material, which upon reaction with water forms, in turn, a C-S-H phase similarly to OPC. Following this method, we produce two types of isochemical cement (referred to as Cel1 and Cel2) by mechanochemical activation of the same hydrothermal batch, and whose chemical and mineralogical compositions are summarized in Table 1. The mechanochemical activation of Cel1 and Cel2 was performed in two different vibrating mills (resp. in PALLA U35 and PALLA VM). Cel1 and Cel2 differ in the turnover of the precursor to hCHS, whose relative content is quantified with a near-infrared (NIR) technique [[17]]. Due to different degrees of transformation during grinding, precursor and resulting hCHS phase are often interlaced in the aggregated grains of the cementitious material.

We prepared standard mortar prisms (4 x 4 x 16 cm$^3$) from both types of cement (after DIN EN 196-1) and determined their compressive strengths to be 16/23/29/31 MPa at 1d/2d/7d/28d, respectively for Cel1 and 24/38/50/50 MPa for Cel2. The larger strength of the sample Cel2 confirms its higher hCHS content. Four grout samples were prepared from each ground material, with water to cement ratio (w/c) of 0.35 (C1, C5) and 0.425 (C3, C6). In the first case, we used 2% superplasticizer (Sky 503, BASF). The corresponding sample notation is presented in Table 1. After 24 hours of hardening, the samples were demolded and stored in water with a constant temperature of 20° C. Eventually, the hydration reaction is stopped by immersing the sample in isopropanol after seven days of hardening.

*1.2. Nano and microindentation*

Before any mechanical testing, the surface of the samples is polished manually by using a series of SiC papers to decrease the surface roughness down to a few hundred nanometers. The nanoscale mechanical properties of the polished samples are then determined by nanoindentation (UNHT, Anton Paar). For each sample, grids of 21x21=441 indentation tests are carried out on a square map using a three-sided, pyramid-like Berkovich indenter. Two subsequent indents are separated by 10 μm, yielding a tested area of 200 μm x 200 μm. Additionally, one sample (C5, Table 1) is tested with a finer grid, in which two subsequent indentations are separated by 5 μm. Each indentation test consists in measuring the indentation depth of the indenter resulting from a symmetric load profile composed of three consecutive steps: a ramp of increasing load at a constant rate (15 mN/min), up to a constant value $P_{max}$= 3 mN. The load is maintained constant at $P_{max}$ for 180 s during which the deformation keeps increasing (creep regime). Finally, the load is decreased linearly to zero, at the same rate as that of the loading step. The maximum load $P_{max}$ yields indentation depths of approximately 300 nm, hence probing the local mechanical properties of the hardened cement paste within a volume of about 1 μm$^3$ [18]. For each indent, the local values of the hardness H and the indentation modulus M are determined from the raw data by the method of Oliver and Pharr [19], while the creep modulus C, which corresponds to the inverse of a creep rate, is computed following the approach developed by Vandamme and Ulm in [8,20]. For a given sample, an indentation grid yields a set of 441 triplets (H, M, C), which is analyzed with a Gaussian Mixture Modeling extensively described in [9]. Such analysis, combined with a Bayesian information criterion, allows us to cluster the mechanical properties and determine the most likely number of phases that the sample is made of at this spatial scale. Finally, we also performed microindentation tests (Microcombi, Anton Paar) following the same protocol and analysis. In that case, the indents are separated by 300 µm and limited to a square map 15x15=225 indents covering a surface of 4.2 mm x 4.2 mm. The (un)loading rate is 15 N/min, and the maximum load is 2 N yielding a typical indentation depth of about 20 µm, which provides coarse-grained information compared to the nanoindentation tests. The holding time is also set to 180 s similarly to the nanoindentation tests, to determine the creep modulus.

*Raman micro-spectroscopy.*

Raman micro-spectroscopy is performed with a spectrograph (Horiba-Jobin-Yvon LabRam Aramis of focal length: 460 mm) equipped with a microscope (Olympus BX41). The excitation wavelength is 473 nm with a power of 11 mW (measured after objective) combined with a grating of 1800 lines/mm. We use a 100x objective (NA=0.9), and the resulting lateral and vertical resolutions are given by the following formulae: $1.22\lambda/NA=0.64$ µm and $4\lambda/NA^2=2.3$ µm, respectively.

The Raman hyperspectral datasets are taken over the same area tested by nanoindentation using a grid with a size of 105 µm x 105 µm with a step size of 2.5 µm for indentations separated by 5 µm resulting in a total of 1764 spectra. For indentations separated by 10 µm, the grid is 210 x 210 µm with a step size of 5 µm. Each spectrum is recorded over 3 s with 12 scans. To image the lateral distribution of a specific phase, within the area measured, several analytical methods can be used: 1) the intensity distribution of characteristic bands, which do not interfere with bands of other phases, 2) The intensity distribution of the whole spectra within the range 200-1200 cm$^{-1}$ (fingerprint range of Raman active normal vibrational modes) using reference spectra of all phases present. For this purpose, we use the Direct Classical Least Squares (DCLS) method implemented in the LabSpec 6 software [21]. Hyperspectral datasets are evaluated based on the principle of superposition of known spectra of the constituents. Such a set of spectra, or and a linear combination of them, form a basis describing only a subspace of the complete vector space. Provided the basis is complete, (i.e., the Raman spectra of all ingredients of the sample are known), it can be used to describe all measured spectra as linear combinations. An example of basis spectra set further used for semiquantitative evaluation is shown in Figure 1.

Thus, every single spectrum of the dataset is fitted using the basis set and corresponding mixing values via least squares method:

$$\vec{S}_i = \hat{B}\,\vec{H}_i + \vec{E}_i$$

$(\vec{S}_i - \hat{B}\,\vec{H}_i)^2 = Min.$ where $\vec{S}_i$ stands for the observed spectrum $i$, $\vec{H}_i$ denotes the mixing values for the spectrum $i$, $\vec{E}_i$ is the error spectrum and $\hat{B}$ denotes the matrix of basis spectra.

Because the Raman spectra measured on such complex systems are always a mixture of different components, it is reasonable to prefer the DCLS procedure. In addition, we use statistical evaluation methods, namely principal component analysis (PCA) and clustering (norm and distance mode: Euclidian, clustering mode: K-means) for comparison purposes.

Finally, areas studied by Raman spectroscopy are also quantified by elemental mapping to compare the phases, and chemical composition found with these two methods, and to map the distribution of alkali elements (Na, K) not precipitating as their own phases.

*2.4. XRD*

X-ray diffraction is performed using MPD Xpert-pro (PANalytical, Almelo, Netherlands) equipped with a multistrip PIXell detector (255 channels, simultaneously covering 3.347° 2Θ) and Cu-radiation. The CuK$_\beta$ is filtered with Ni-filter. Measurements are conducted with adjustable slits giving a constant irradiated sample length of 10 mm and soler slits 0.04 rad (2.3°). Phase identification is performed with the software packages Highscore-Plus (PANalytical) and Diffrac-Plus (Bruker-AXS, Karlsruhe, Germany). Bulk quantitative phase analyses of Cel1 and Cel2 with external standard (20 wt% α-$Al_2O_3$) are performed following the Rietveld method with TOPAS V4.2 (Bruker-AXS).

*2.5. Scanning Electron Microscopy and Elemental Mapping:*

Samples are imaged with a Scanning Electron Microscope (SEM, Zeiss Supra 55VP) with a field emission gun operating at 20 kV accelerating voltage. An aperture of 30 μm was used for imaging, mostly via secondary electrons (mode SE2) and in some cases also via backscattered electrons (mode AsB). Elemental mapping of the indentation areas is performed using a Quantax EDS system (XFlash 5010 with 10 mm$^2$ detection area and 127 eV FWHM/MnK$\alpha$) coupled to the SEM.

2. **RESULTS AND DISCUSSION**

*3.1. Chemical and phase composition*

In the present study, one of the most challenging tasks is the determination of the phase composition of Celitement samples because both the hydrothermal precursor (C-S-H) and the hydraulic product (hCHS) are X-ray amorphous [22]. Additional complication results from the presence of gel water, in addition to adsorbed and chemically bound water, amorphous $CaCO_3$ and $Ca(OH)_2$, which quantity should be determined by thermogravimetry. The results of the quantitative analyses with the Rietveld methods are shown in Table 1. The hydrothermal precursor used for mechanochemical activation consists of 83.5 wt.% of amorphous product, the main component being C-S-H phase of type C-S-H I, but both amorphous $CaCO_3$ (ACC) and $Ca(OH)_2$ were also present. Furthermore, some trace of crystalline phases are also present: quartz, feldspars, and mica from the industrial-grade quartz sand and portlandite, spurrite, and $C_2S$ from the slaked lime used. Moreover, some $\alpha$-$C_2SH$ and 11Å tobermorite are formed as by-products of the hydrothermal treatment. Grinding the sample results in a higher quantity of the amorphous material (about 89 wt.%), and a reduction of the $Ca(OH)_2$ and calcite. Note that tobermorite and mica are not detectable. Generally, the ground samples from both mills show a similar phase composition with a slight tendency for a higher amount of amorphous content with increasing grinding intensity for Cel2. Finally, table 1 also provides results from XRF analysis showing a C/S ratio of 1.28 and loss on ignition (LOI) of 11.4 %. Thermal studies show that the LOI is resulting mainly from a loss of $H_2O$ (7.62-7.93 wt.%) and $CO_2$ (3.53-3.30 wt.%). Cel1 and Cel2 differ in the amount of hydraulically reactive component, hCHS, being 36 wt.% and 45 wt.%, respectively.

Table 1: Chemical and mineralogical composition of isochemical cements measured with XRD, XRF, and TA (LOI) in wt.% and the composition of lime samples C1, C3, C5, and C6.

| Mineralogical composition | | | Chemical composition | |
|---|---|---|---|---|
| XRD | Cel1 (wt. %) | Cel2 (wt. %) | XRF | (wt. %) |
| amorphous | 88,6 | 89,6 | $SiO_2$ | 42,28 |
| α-$C_2SH$ | 0,9 | 0,7 | $Al_2O_3$ | 2,73 |
| Quartz | 5,3 | 4,7 | $Fe_2O_3$ | 0,67 |
| Portlandite | 1,4 | 0,8 | CaO | 50,33 |
| Calcite | 1,2 | 1,0 | MgO | 1,04 |
| β-$C_2S$ | 0,1 | 0,6 | $SO_3$ | 0,13 |
| Spurrite | 1,3 | 1,2 | $K_2O$ | 1,22 |
| Feldspars | 1,2 | 1,3 | $Na_2O$ | 0,24 |
| | | | others[1] | 0,19 |
| Σ | 100,0 | 100,0 | Σ | 98,82 |
| | | | C/S | 1,28 |
| TA | | | LOI 950°C | 11,39 |
| $H_2O$[2] | 7,62 | 7,93 | | |
| $CO_2$ | 3,53 | 3,30 | | |
| ACC[3] | 6,70 | 6,39 | | |
| grout samples | | hCHS[4] (wt. %) | w/c | Sky 503 |
| C1 | 20 g Cel1 | 36 | 0,350 | 2,00 |
| C3 | 20 g Cel1 | 36 | 0,425 | |
| C5 | 20 g Cel2 | 45 | 0,350 | 2,00 |
| C6 | 20 g Cel2 | 45 | 0,425 | |

[1] others include $TiO_2$, $Mn_2O_3$, and Cl

[2] $H_2O$ was determined by weight loss between 100-570° C, $CO_2$ between 570-1000° C

[3] ACC = amorphous calcium carbonate determined by TA [total $CO_2$ loss – contribution of spurrite (10 %) → total $CaCO_3$ – calcite (XRD)]

[4] hCHS was determined by NIR

*3.2. Raman basis spectra.*

The interpretation of the data obtained on hydrated Celitement pastes, by Raman micro-spectroscopy, necessitates identifying the phases present by their characteristic spectra. Figure 1 shows the typical basis spectra for reacted Celitement samples consisting of quartz, feldspar, anatase (sand), spurrite, β-$C_2S$ (slaked lime), silicon carbide (from polishing) and C-S-H phase (hydration product). Carbonation of fresh C-S-H leads to the formation of $CaCO_3$, mostly calcite. The polymorphic modifications of $CaCO_3$ calcite, aragonite, vaterite, and even amorphous $CaCO_3$ can be unambiguously differentiated [23]. However, for simplicity, we refer to all of them as $CaCO_3$ in this manuscript. Raman spectra of synthetic C-S-H phases with different C/S ratios have been extensively studied in [10,11]. While C-S-H was readily detected in the Raman spectra of $C_3S$ pastes [24], it remains a challenging task in OPC pastes [25] for which only a few studies have detected C-S-H unambiguously [26]. The primary difficulty is due to the very complex nature of OPC and its hydration products. Second, the low degree of crystallinity and disordered nature of C-S-H compared to other hydration products like CH, ettringite, monosulphate, etc. and unreacted clinker minerals contributes to this problem. Studying Celitement pastes presents a unique opportunity not only to detect and image the distribution of C-S-H, but also to differentiate between C-S-H with different density in terms of high density (HD) and low density (LD) C-S-H [1]. The presence in cement pastes of regions with different density is undisputed, although several theories of their origin and development exist as summarized in [4]. The principal difference between LD and HD C-S-H is the presence of gel pores filled with water in the former. Therefore, if one were able to image the distribution of the gel pores this would reflect the areas with LD C-S-H. Such an approach was followed in [4,27] by using H NMR. Here we use Raman micro-spectroscopy to monitor alcohol, which can enter the permeable LD C-S-H structure and exchange water in capillary and gel pores. Alcohol is widely used to stop hydration after a certain time interval. Here we use isopropanol, which can be readily observed in Raman spectra by its band at 820 $cm^{-1}$ that always appears together with the well-known C-S-H bands at 670 $cm^{-1}$ and between 800-1000 $cm^{-1}$ (very broad). Thus, the intensity of the isopropanol band is used as the main signature to differ between HD and LD C-S-H. In addition, there are differences in the spectral range 800-1000 cm$^{-1}$ between both C-S-H most probably due to different degrees of silicate polymerization [11]. The spectra of LD and HD C-S-H show low levels of carbonation seen by the broad band between 1050-1100 cm$^{-1}$ typical for amorphous $CaCO_3$ [23]. Sharper bands of calcite are superimposed to this broad band.

Another approach to image areas with LD C-S-H consists of using the carbonate bands, suggesting a strong carbonation process. Indeed, humidity and open porosity accelerate carbonation along with other physicochemical factors like $CO_2$ concentration, temperature, pressure, and pH [28]. Therefore, by keeping the latter constant, it is evident that LD C-S-H would be more prone to carbonation than HD C-S-H. Indeed, we observe that fresh pastes (with a higher degree of capillary and gel porosity) are more likely to carbonate than hardened pastes (with a substantially lower degree of porosity) [29].

Moreover, one of the side effects of carbonation is the reduction of the sample porosity [30]. Carbonation causes a change of the pore size distribution reducing the gel porosity (i.e., the porosity accessible to $H_2O$) by clogging with $CaCO_3$ and increasing the capillary porosity [29,31]. As a consequence, the permeability of the sample for water is also reduced [32]. Therefore, some areas of LD C-S-H, which underwent strong carbonation may not be accessible (or only partially) for isopropanol. That is the reason why a combination of different markers should be used for unambiguously imaging LD C-S-H.

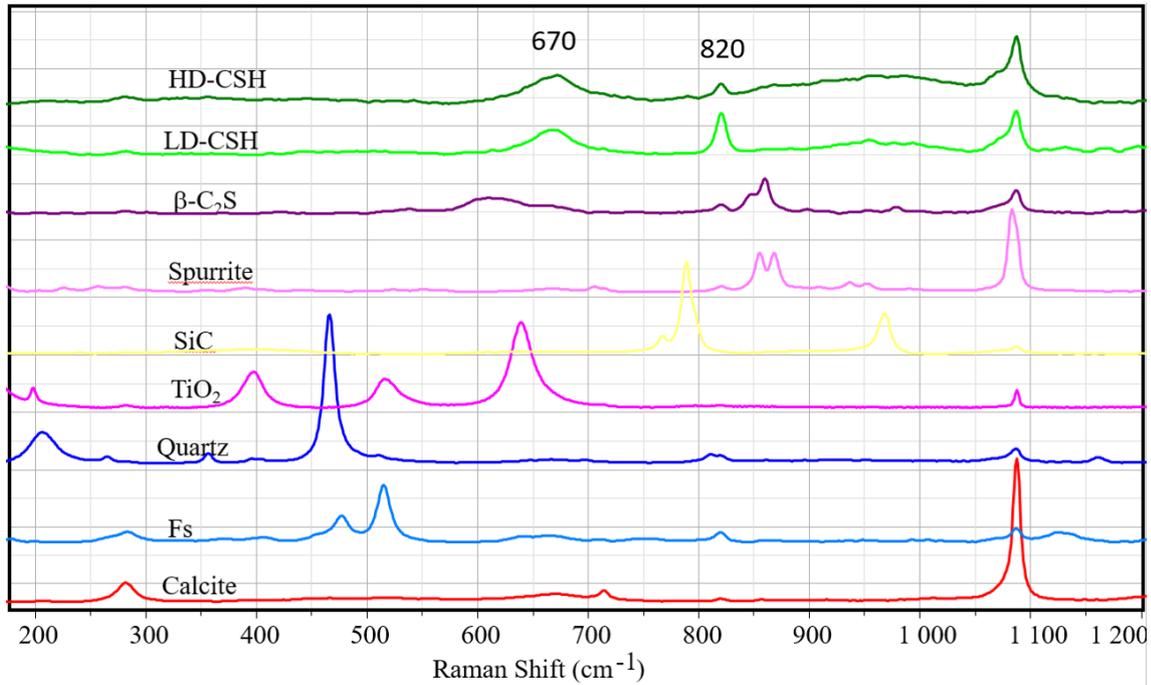

**Figure 1:** Basis set of Raman spectra of the constituents of hardened Celitement paste used for the evaluation of sample C3 by the DCLS method.

*3.3. Microindentation*

The micromechanical properties of four different samples (C1, C3, C5, C6) are reported in Figure 2. The data consist of master curves in various representations (M vs. H, M vs. C, etc.), which show a consistent trend with the formulations and measured strengths. The weakest mechanical properties are measured for sample C3, which has the lowest content of hydraulic phase and highest w/c ratio. With increasing cement content or decreasing w/c ratio, the mechanical properties improve. Therefore, it seems possible to adjust the mechanical properties of the samples not only by varying the ratio w/c, but also by tuning the content of hCHS. Note that similar results were already obtained for hardened paste of OPC [33].

The results from microindentation prompt to take a closer look at the both samples with most extreme mechanical properties: i.e., C3 with the lowest hCHS content and the highest w/c ratio, and sample C5 with the highest hCHS content and lowest w/c ratio.

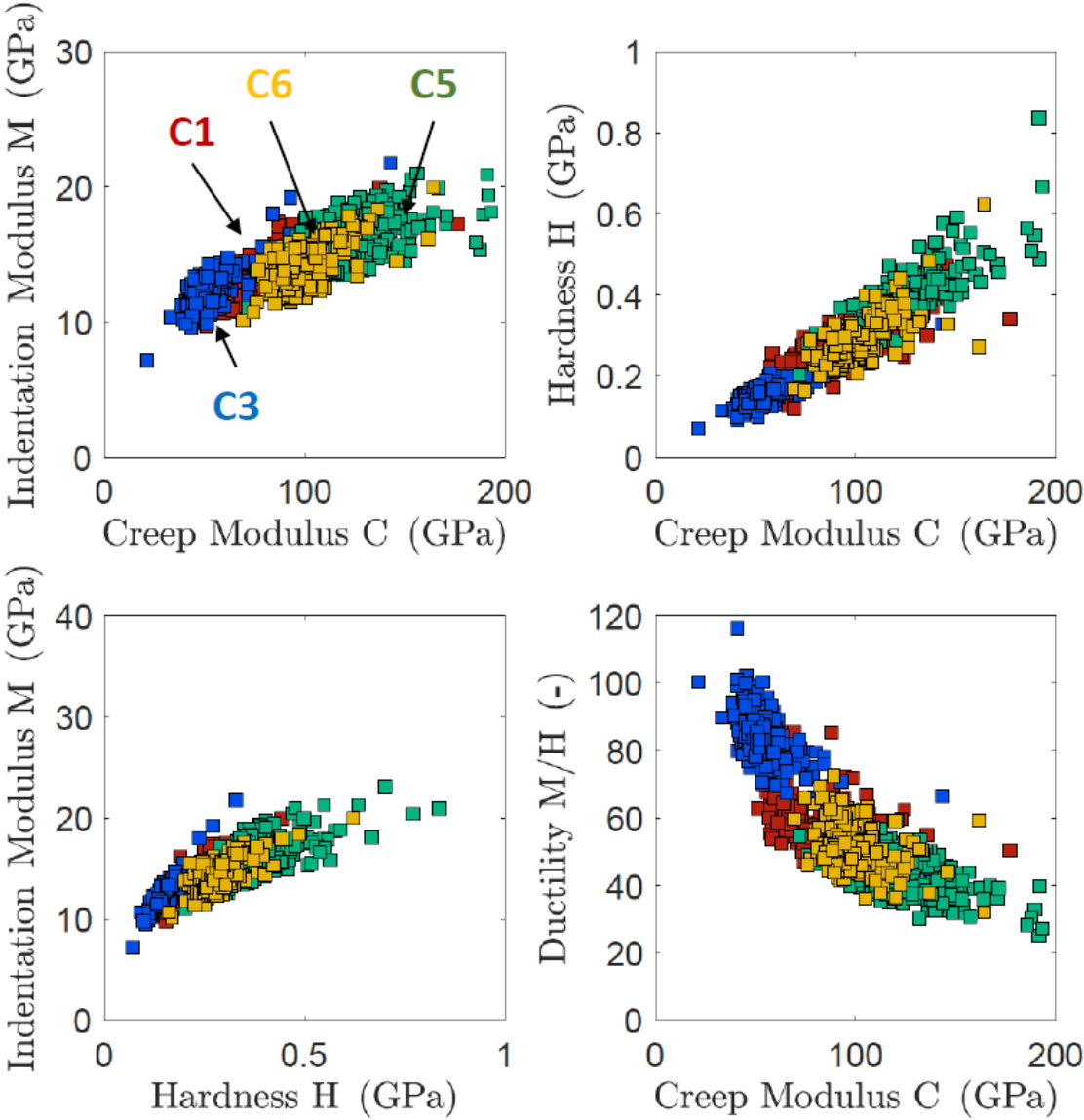

**Figure 2:** Micromechanical properties H, M, and C of samples C1, C3, C5, and C6. The data are plotted in various ways, and for all graphs, the data collapse on a master curve. The data set for each sample corresponds to an indentation grid of 15x15=225 indents spreading over an area of 4.2 mm x 4.2 mm.

*3.4. Sample C3*

The sample C3 is tested by nanoindentation, and figure 3 shows an optical photograph of the indentation grid 210 x 210 μm (Figure 3A). The dark region visible on the upper right side of the image corresponds to an area that has been damaged during the polishing step. Figure 3B-3F show maps of the distribution of the following elements: Ca (B), Si (C), K (E), Na (F) and an SE image of the same area (D). The cracks observed in the SE image result from the partial dehydration of the sample under high vacuum conditions in the electron microscope. The distribution of Ca is almost uniform, except for dark areas belonging to quartz and feldspar particles. The silicon distribution shows clearly euhedral particles of quartz in a siliceous matrix. The sodium map helps to identify a euhedral feldspar particle of albite type. Finally, potassium seems to be distributed unequally: 1) as small euhedral particles (K-feldspars) and 2) in a matrix surrounding dark K-free areas.

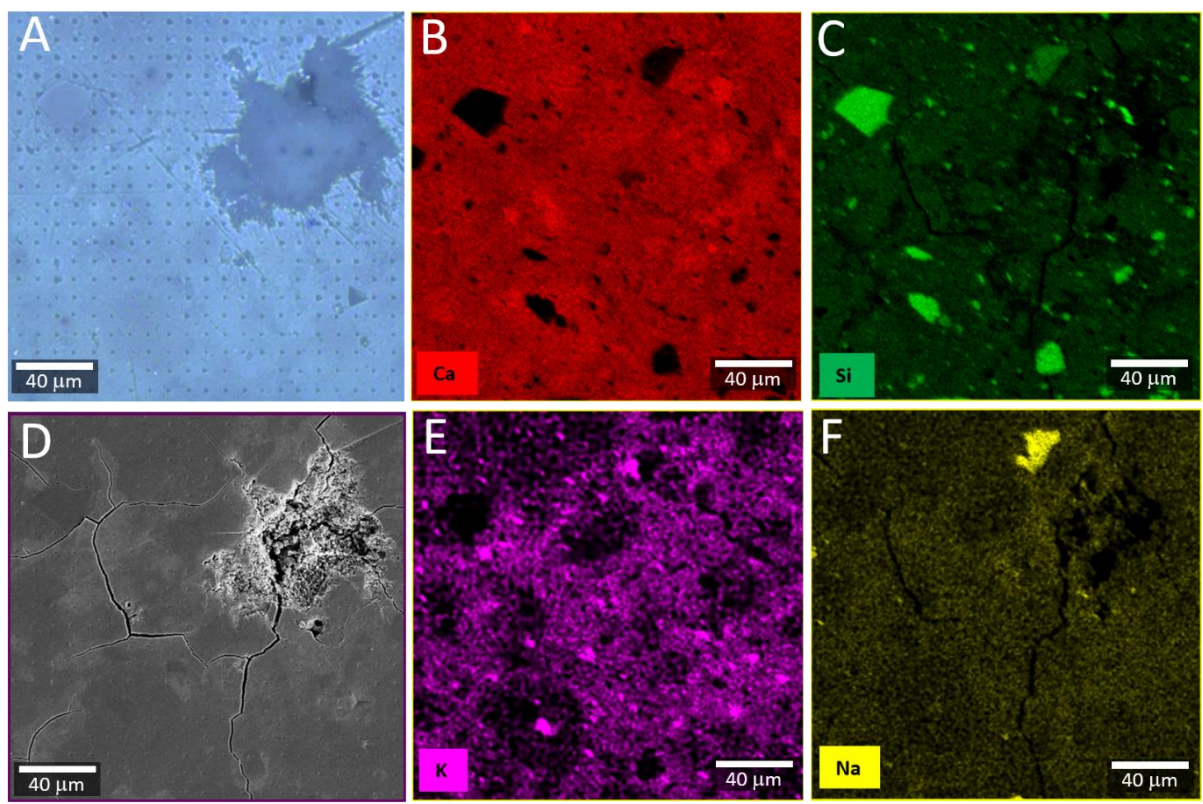

**Figure 3:** A: Optical image of the nanoindentation grid performed on sample C3. B, C, E, F: Images of the elemental distribution of Ca, Si, K, and Na. D: an electron microphotograph of the same area

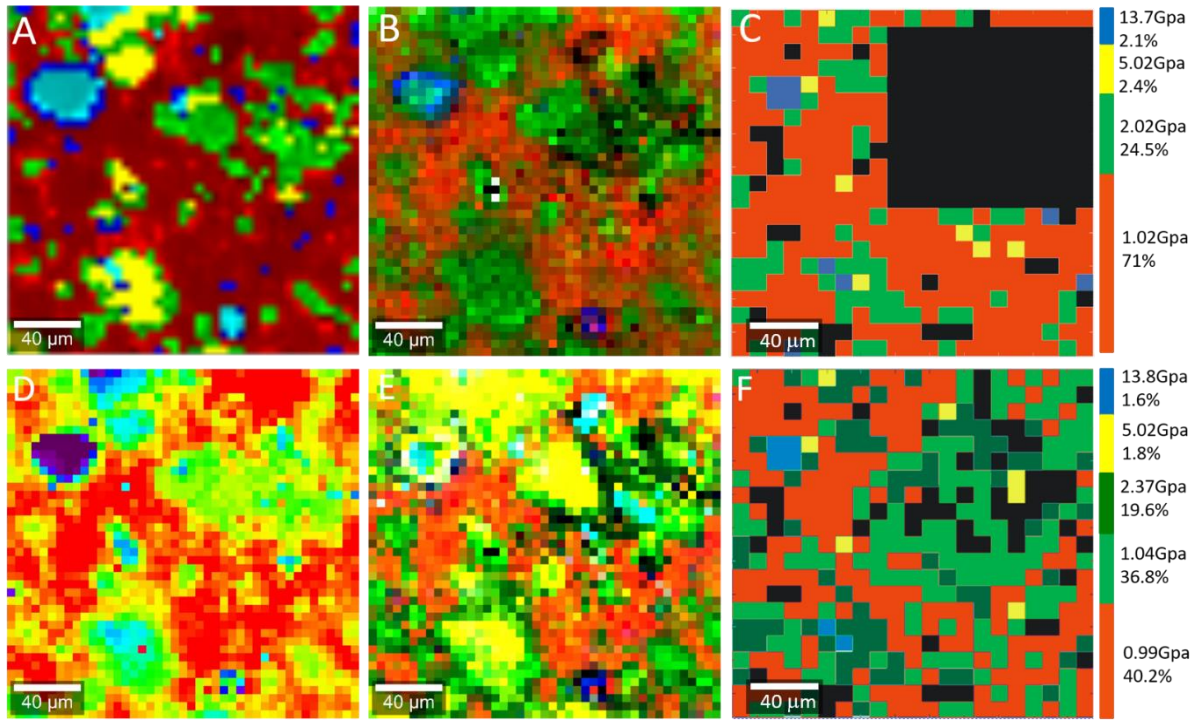

**Figure 4**: A, B, D, and E Phase distribution in sample C3 revealed by different evaluation methods of the results from Raman mappings. A: results from clustering (6 clusters) without prior information about the phase composition. D: Results from PCA with assigning the spectra of 6 principal components. B: Results from imaging of the lateral distribution of the specific bands for quartz (462 cm$^{-1}$, blue), carbonate (1086 cm$^{-1}$, red) and C-S-H (670 cm$^{-1}$, green). E: same as in B complemented with feldspar (520 cm$^{-1}$, light blue) and isopropanol (820 cm$^{-1}$, yellow). C and F: Results from the clustering of the nanoindentation data with a Gaussian Mixture Modeling method with corresponding values for hardness. Black areas are "bad pixels" where the data could not be analyzed. In C, the whole area damaged during polishing was removed before clustering the remaining data.

The evaluation of the results of the Raman mappings of sample C3 is presented in Figure 4A, 4B, 4D, and 4E along with results from the clustering of the nanoscale mechanical properties by Gaussian Mixture Modeling coupled to a Bayesian criterion. Comparison between clustering (A), PCA, and lateral distribution of the specific bands for the main phases present, namely quartz (462 cm$^{-1}$), carbonate (1086 cm$^{-1}$) and C-S-H (670 cm$^{-1}$) (D). In Fig. 4E the phases are complemented with feldspar (520 cm$^{-1}$) and isopropanol (820 cm$^{-1}$).

All Raman images in Fig. 4 show a similar pattern with different degrees of detail, pointing to the presence of aggregates in a matrix of carbonated material. The latter is always associated with C-S-H (i.e., its Raman spectra always display C-S-H bands) and thus could be regarded as freshly formed LD C-S-H. Areas with a lower degree of carbonation and higher C-S-H content could be assigned to former autoclave C-S-H in the vicinity of which HD C-S-H is formed upon hydration. The presence of C-S-H spectra with clearly pronounced bands corresponding to isopropanol (820 cm$^{-1}$), which was used to stop the hydraulic reaction in the samples, is very informative as it substitutes for $H_2O$ in the pores, its distribution can be used as a marker for porosity. Spectra of C-S-H without or with a lower amount of isopropanol bands can be regarded as belonging to HD C-S-H (Fig. 1). The results of multivariate GMM clustering of the nanomechanical properties are presented in Fig. 4C and 4F. In Fig. 4C the nanoindentation results from the damaged area ( see Fig. 3A, 3D) are excluded from the clustering process. The remaining indentations show about 70 % values for hardness around 1 GPa. Their distribution coincides almost perfectly with the distribution of LD C-S-H from the Raman maps. About 25 % indents show an average hardness of about 2 GPa, which coincides with the HD C-S-H areas in the Raman images. The remaining 2.5 % indentations show considerably higher hardness (5 GPa and 13.8 GPa), coinciding with feldspars and quartz in the Raman maps. In Figure 4F the damaged area was not excluded from evaluation allowing small differentiation within the LD C-S-H in two ranges according to their hardness (0.99 GPa and 1.04 GPa). This evaluation shows even better coincidence with the Raman maps. In both cases, the area of the LD-C-S-H is >70 % (71 %, 77 %, respectively), which is due to the high w/c ratio of sample C3.

Finally, the comparison of the distribution of alkali elements (especially K) with the Raman maps, clearly shows enrichment in the LD C-S-H.

*3.5. Sample C5.*

Figure 5 shows the result of the clustering analysis of the nanomechanical data yielding four clusters with different values for hardness (H), indentation modulus (M), and creep modulus (C). Clusters 1 and 3 represent nearly 94% of the measured area and correspond to two different types of C-S-H. The higher values (in GPa) for H (1.80), M (36.8) and C (382) for cluster 1 compared to cluster 3 (H: 1.36, M: 28.7, C: 298) imply that the former represents the areas of HD C-S-H and the latter these of LD C-S-H. Clusters 2 and 4 correspond to particularly dense structures (especially cluster 4), which represents probably accessory minerals like quartz and feldspar. In contrast with sample C3, the distribution of LD and HD C-S-H is different. While sample C3 shows 71-77 % LD C-S-H and only 20-24 % HD-C-S-H, sample C5 displays about 52 % LD C-S-H and 43 % HD C-S-H. Additionally, the values for hardness (especially for LD C-S-H, 1.36 GPa) are much higher than those of sample C3 (about 1 GPa). This result is due to two factors: a higher content of reactive material (hCSH) and a lower w/c ratio of sample C5.

Raman maps of sample C5 created by the intensity distribution of particular bands of C-S-H (670 cm$^{-1}$) [11], $CaCO_3$ (1070-1090 cm$^{-1}$), quartz, and feldspar (440-540 cm$^{-1}$) or isopropanol (820 cm$^{-1}$) are shown in Fig. 6A and 6C. Once again, the presence of isopropanol bands gives valuable information about the porosity of the microstructure. A comparison with the lateral distribution of the GMM clusters points to an excellent agreement between the distribution of HD C-S-H from Raman map and cluster 1 (Fig. 5). Correspondingly, areas of C-S-H with high isopropanol distribution (LD C-S-H) coincide with cluster 3. Carbonated areas show rather mixed behavior of the mechanical properties pointing to possible consolidation due to calcite formation. In these areas, fewer isopropanol bands are seen, suggesting a lower porosity.

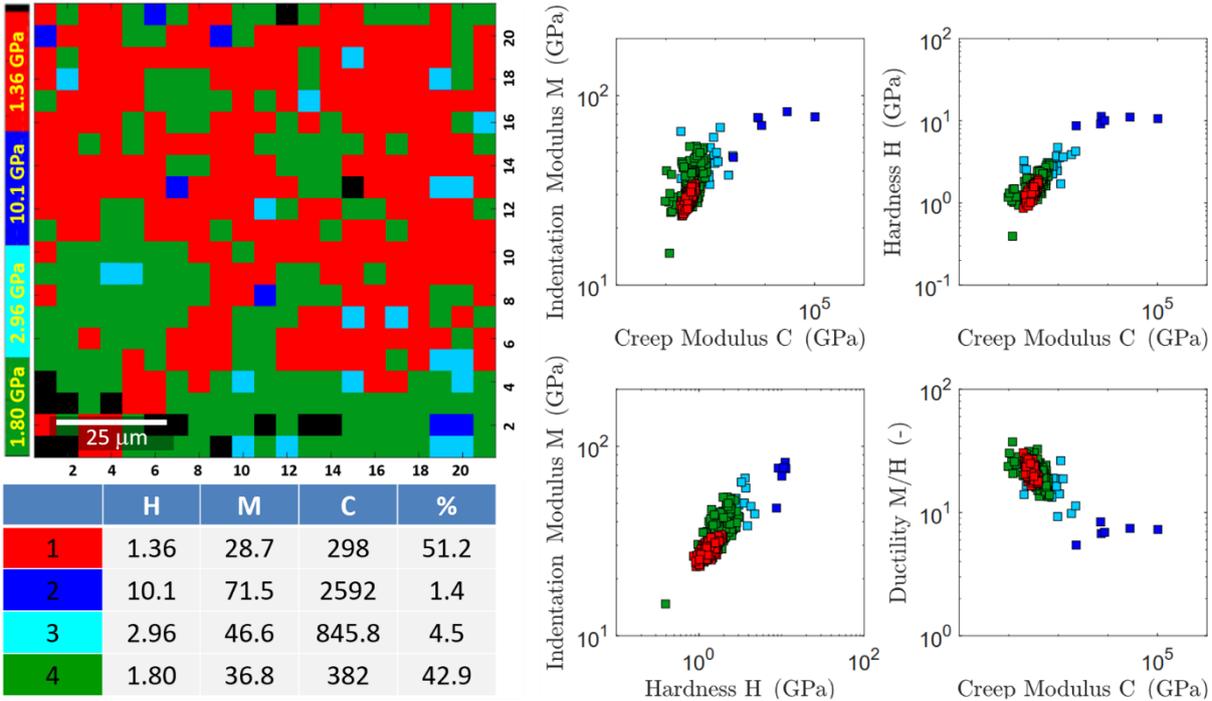

**Figure 5**: Results from the nanomechanical testing of sample C5 The clustering of the data (Hardness H, Indentation Modulus M, and Creep C) yields 4 clusters. Their lateral distribution is displayed on the map on the top left side. Black areas (14 points) are outliers that cannot be analyzed.

Maps of the phase distribution after DCLS (Fig. 6E and 6F) show much more detailed phase information compared to those in Fig. 6A and 6C. They present an excellent coincidence with the results from clustering indentation data and Raman images (Fig. 6C, 6F). Aggregates seen by optical, EDX, and Raman images correspond to denser areas, which consist of HD C-S-H. Residual $Ca(OH)_2$ from cement processing is always located next to HD C-S-H and intercalate neither with the freshly formed C-S-H, nor with the carbonated areas. HD C-S-H generally shows a higher intensity and a lower Full Width at Half Maximum (FWHM) of the 670 cm$^{-1}$ band (Si-O-Si) than LD C-S-H. However, some broadening at higher frequencies is observed due to compressive stress arising from C-S-H growing in confined space. The availability of precipitation space seems to be the main factor for the densification of the cement matrix, according to recent research [3,34].

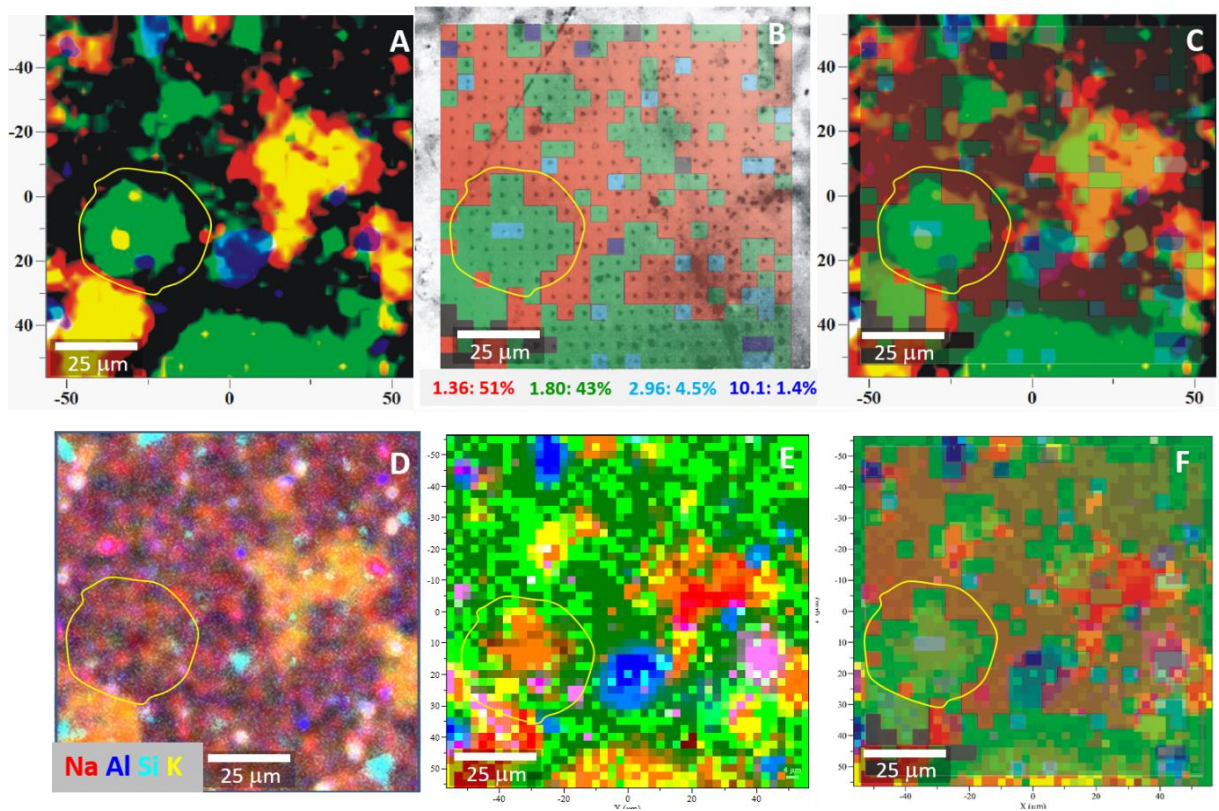

**Figure 6**: A: Raman map of the distribution of C-S-H (green), carbonate (red), quartz, and feldspars (blue) of sample C5. The black areas correspond to C-S-H with high porosity (LD C-S-H). B: Cluster distribution map of the mechanical properties (H, M, and C) superimposed on the optical image. C: Superimposed images of the Raman map in A with the cluster distribution map of the H, M, and C. D: Elemental distribution (EDX). E: Phase distribution after DCLS: Ca(CO$_3$) (red), quartz and feldspar (blue), anatase (magenta), rutile (brown), HD C-S-H (dark green), LD C-S-H (light green), CH (orange).

F: Cluster distribution map of the mechanical properties (H, M, and C) superimposed on the image in E. In all six images, the yellow circle indicates the contours of an aggregate.

EDX map of the same area (Fig. 6D) helps to understand the fate of alkali elements, which seem to be mobilized during hydration, as their distribution coincides with that of LD C-S-H. Increased carbonation is observed only in LD C-S-H areas due to higher amounts of $CO_2$ in the solution. Isopropanol can be used as a marker for detecting the presence of large capillary pores accessible to $H_2O$, thus being observed in higher concentrations in the LD C-S-H.

*3.6. Sample C6*

Shadowy outlines of original cement grains and secondary minerals are visible in the optical image of the polished surface of sample C6 displayed in Fig. 7A. A newly formed matrix stands out brighter due to its high surface area and the resulting increased scattering. Scratches from the polishing are also visible. A Raman image of all types of C-S-H is shown in Fig. 7B. One can see clear differences in intensity between dense, primary grains and newly formed LD C-S-H, while black areas correspond to accessory minerals like quartz and feldspars from sand and calcite formed by carbonation.

Raman mapping of a band corresponding to isopropanol that arises from preparation shows ring-shaped accumulations around primary cement grains (Fig. 7D). This zone is interpreted as precipitated semipermeable calcium-rich C-S-H gel with high, open porosity, which is not interlinked with the pore solution.

In Raman spectra, crystallographic stress in minerals tends to broaden and shift single vibrations. Figure 7E shows stressed C-S-H in the center of the former cement grains, corresponding to HD C-S-H. The porosity of LD C-S-H formed from the solution is mainly closed due to carbonation (Fig. 7F). LD C-S-H also presents some isopropanol but with reduced intensity. Calcite crystals occur only sporadically, as a primary impurity.

By mapping the same area with EDX, areas of the sample grown from the pore solution can be distinguished from areas converted by diffusion of water and non-hydraulic accessories. A mapping of the potassium concentration is shown as an example. Highly mobile potassium is enriched exclusively in the area of the original pore solution. The distribution pattern corresponds to that of LD C-S-H. High potassium concentrations in sharply bordered grains correspond to potassium feldspar from the raw materials.

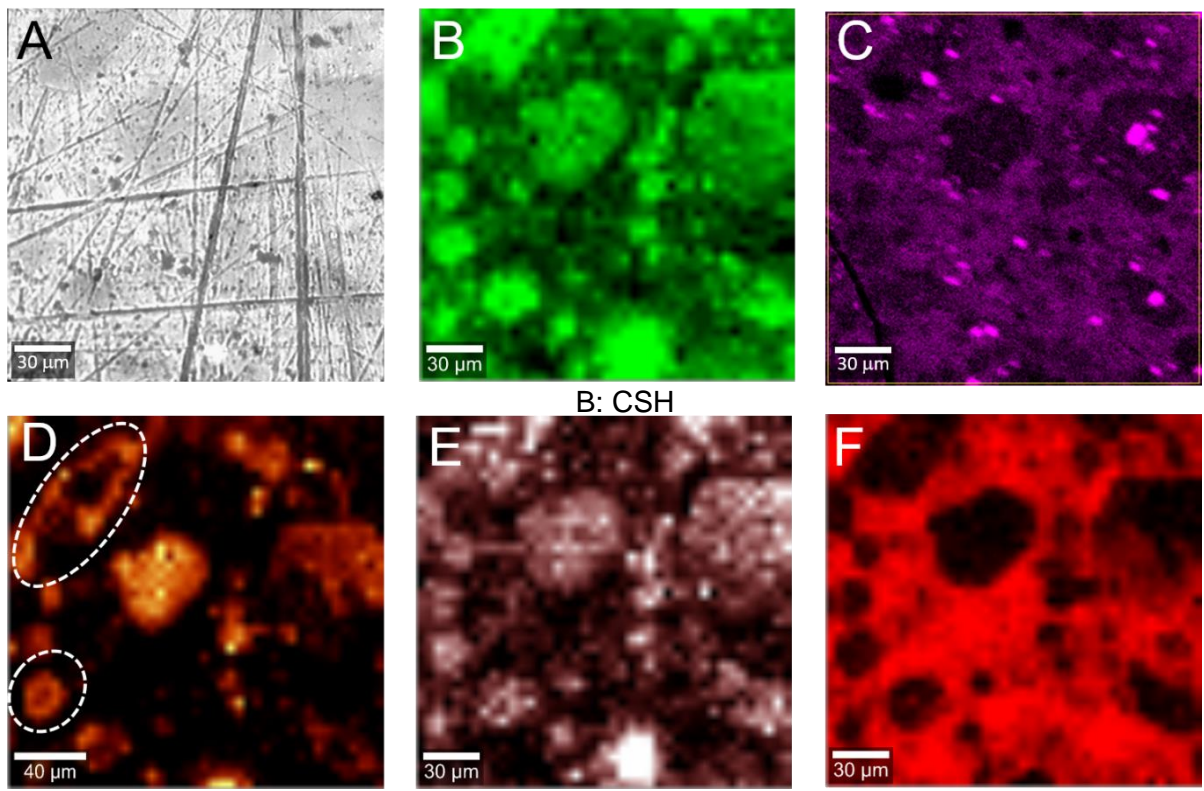

**Figure 7:** Images of the indented area of sample C6. A: Optical image; B: Raman map of the distribution of C-S-H (green, LD C-S-H + HD C-S-H); C: Elemental mapping of potassium (purple, EDX); D: Raman map of the distribution of isopropanol, labeling precipitated calcium-rich C-S-H gel; E: Raman map of the distribution of HD-CSH labeled by stress broadening of vibrations; F: Raman map of the distribution of carbonate.

By clustering the nanomechanical measurements from sample C6 with GMM, three clusters were identified and assigned to LD C-S-H (red), HD C-S-H (green), and accessories (blue) (see Fig. 8). An illustration of the distribution of these clusters in the inspected area is shown in Figure 8 (left). Figure 8 (right) shows the same area with the same color-coding from a combination of Raman data. There is an excellent agreement between both maps, with a visibly higher resolution of the Raman measurements. The majority of the nanoindentation values correspond to a paragenesis of phases. The percentage of the distribution of the LD (68.9 %) and HD C-S-H (27.6 %) is somewhat similar to that of sample C3, which is related to the same w/c ratio of both samples. Nevertheless, sample 6 shows not only higher content of HD C-S-H, but also higher values for hardness due to the higher content of hCSH.

The course of hydration of sample Cel2 with a w/c of 0.425 can be described qualitatively from the measured maps. After being mixed with water, fine cement grains (< approx. 5μm) are completely dissolved. For larger grains, the dissolution is limited to marginal areas. The pH increase associated with the progressing dissolution of cement is significantly slower than in OPC. Therefore, well-ordered LD C-S-H phases with a low C/S ratio are formed, which explains the strong fixation of alkalis such as potassium to LD C-S-H. The constant dissolution of hCHS and precipitation of LD C-S-H with a low C/S ratio increase the ionic strength of the solution, which finally leads to the precipitation of a silicate-rich semi-permeable C-S-H gel around large remaining cement grains. This new interface is permeable for water. Thus, the subsequent hydration of the larger grains is controlled by diffusion through the C-S-H barrier, which drastically slows down the reaction kinetics. Water, driven by osmotic pressure, passes the gel layer into the interior of the grain. The limited space for hydration inside the grain leads to the formation of stressed HD C-S-H. Similar reactions are described for $C_3S$ [35] and are consistent with the results of Königsberger et al [34].

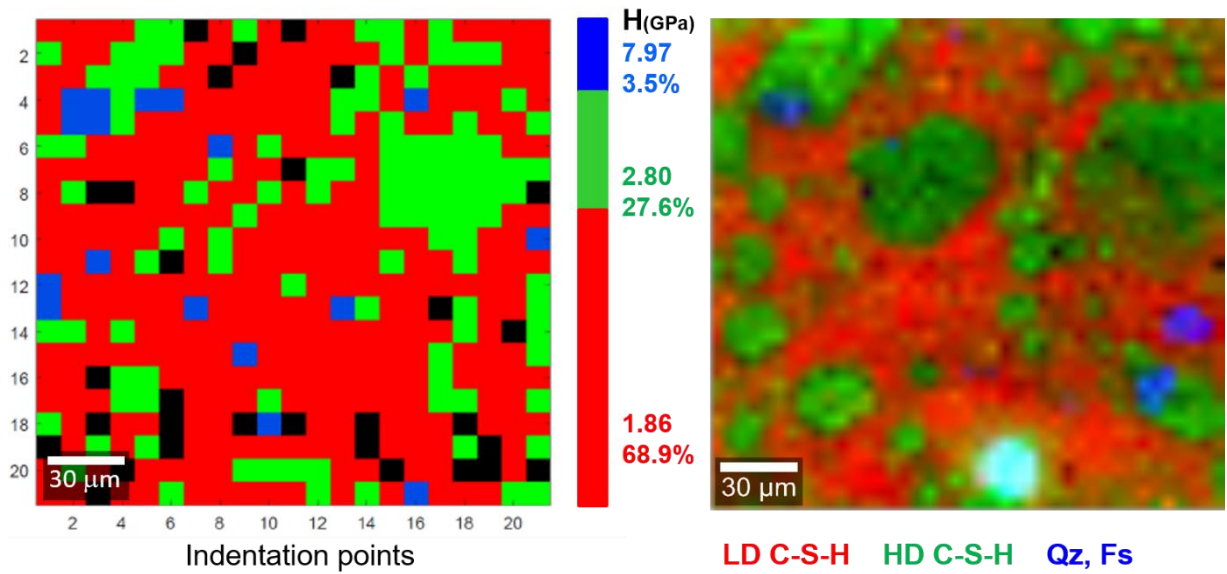

**Figure 8: (left)** GMM clustering of the nanomechanical data (H, M, and C) yields 3 clusters LD-C-S-H (red, 68.9%), HD-C-S-H (green, 27.6 %) and accessories (blue, 3.5 %). The map corresponds to a square indentation grid composed of 21 x 21=441 indents separated by 10 μm. **Right**: the same area with the same color code from a combination of Raman data.

## 3. CONCLUSIONS

We have shown that Raman micro-spectroscopy is a valuable tool for representing the lateral phase distribution in hardened cementitious materials. Raman micro-spectroscopy can be used for assigning unambiguously specific mechanical properties to a particular phase. Comparatively, PCA, clustering, DCLS, or direct imaging of specific Raman bands deliver similar results. Furthermore, we have shown the possibility to image the distribution of different C-S-H phases using the bands of isopropanol and carbonate as a probe for porosity.

Linking the presence of HD C-S-H to areas with confined space proves recent ideas of the development of the microstructure of hardened cement paste.

While Raman micro-spectroscopy has a very good lateral resolution (in the order of magnitude of the excitation wavelength or better), the real limiting factor is the resolution of the nanoindentation grid. Because the measured Raman spectra are mostly a result of the superposition of the spectra of individual components mixed on a very fine scale, a quantitative approach on every point would be very helpful in order to reveal the exact composition at each locus of indentation. In order to take into account polarization and orientation effects on the Raman spectra we need to use chemometric methods or exact determination of the Raman cross-sections of all phases to obtain real quantitative information at every measurement point.

Isochemical cement ("Celitement") consists of calcium hydrosilicates, which in contact with water, react to form C-S-H phases similar to those already known from the hydration of the clinker minerals $C_3S$ and $C_2S$. The production process requires considerably less lime than OPC. In particular, the absence of $Ca(OH)_2$ as a hydration product makes it possible to change both the composition and the mechanical properties of the hydrated materials.

In this work, we have imaged hydrated samples of isochemical cement of different w/c values with Raman micro-spectroscopy. Their characteristic Raman spectra were used to determine the distribution of hydration products HD C-S-H and LD C-S-H and small amounts of calcium carbonate. Furthermore, non-reactive secondary components such as quartz and feldspar were identified.

The reaction zones as well as the occurrence of HD and LD C-S-H can be explained qualitatively with a simple reaction model, which requires the precipitation of a semipermeable C-S-H gel layer around larger cement grains during the reaction. Mechanical properties derived from the clustering of nanoindentation measurements with a lateral resolution of 5 micrometers agree very well with the observed model. However, the resolution of nanoindentation measurements is not high enough to use accurate micromechanical models. The simple reaction chemistry of this type of cement allows tailoring their microstructure by varying their composition and adding buffering systems for control of the reaction kinetics.

The results show a meaningful trend of development of both the mechanical properties and the quantitative distribution of LD and HD C-S-H in relation with hC-H-S quantity and w/c ratio. Nevertheless, we should be aware of the fact that the results are only a snapshot (after 7 days of hydration) of a dynamic system, which needs further research and understanding.